\documentclass[12pt, draftcls,onecolumn]{IEEEtran}

\usepackage{cite}
\usepackage{amsmath,amssymb,amsfonts}
\usepackage{graphicx}
\usepackage{epsfig}
\usepackage{color}
\usepackage[caption=false]{subfig}
\usepackage{lettrine}
\usepackage{float}
\usepackage{bbm}
\usepackage{lipsum}
\usepackage{bm}
\usepackage{mathtools}

\usepackage{tikz,xcolor,hyperref}

\definecolor{lime}{HTML}{A6CE39}
\DeclareRobustCommand{\orcidicon}{
	\begin{tikzpicture}
	\draw[lime, fill=lime] (0,0) 
	circle [radius=0.16] 
	node[white] {{\fontfamily{qag}\selectfont \tiny ID}};
	\draw[white, fill=white] (-0.0625,0.095) 
	circle [radius=0.007];
	\end{tikzpicture}
	\hspace{-2mm}
}

\foreach \x in {A, ..., Z}{\expandafter\xdef\csname orcid\x\endcsname{\noexpand\href{https://orcid.org/\csname orcidauthor\x\endcsname}
			{\noexpand\orcidicon}}
}

\def \treq {\stackrel{\tiny \Delta}{=}}

\newcommand{\E}{\ensuremath{\mathbb E}}
\newcommand{\done}{\hfill $\blacksquare$}
\newcommand{\ra}{$\mathcal{R}_1$}
\newcommand{\rb}{$\mathcal{R}_2$}
\newcommand{\src}{$\mathcal{S}$}
\newcommand{\des}{$\mathcal{D}$}
\newtheorem{proof}{Proof}
\newtheorem{lemma}{Lemma}

\newtheorem{proposition}{Proposition}

\makeatletter
\def\@seccntformat#1{\@ifundefined{#1@cntformat}%
	{\csname the#1\endcsname\quad}
	{\csname #1@cntformat\endcsname}
	}
\makeatother

\date{}
\begin{document}

\title{Can a Multi-Hop Link Relying on Untrusted Amplify-and-Forward Relays Render Security?}



\author{
	
\IEEEauthorblockN{
Milad Tatar Mamaghani\IEEEauthorrefmark{1}, Ali Kuhestani\IEEEauthorrefmark{2}, and Hamid Behroozi\IEEEauthorrefmark{2}}
		
\IEEEauthorblockA{
\IEEEauthorrefmark{1}Electrical and Computer Systems Engineering Department, Faculty of Engineering, Monash University, Melbourne, Australia\\
\IEEEauthorrefmark{2}Electrical Engineering Department, Sharif University of Technology, Tehran, Iran}}

\maketitle
\markboth{}{}

\begin{abstract}
Cooperative relaying is utilized as an efficient method for data communication in wireless sensor networks and the Internet of Things. However, sometimes due to the necessity of multi-hop relaying in such communication networks, it is challenging to guarantee the secrecy of cooperative transmissions when the relays may themselves be eavesdroppers, i.e., we may face with the untrusted relaying scenario where the
relays are both necessary helpers and potential adversary. To obviate this issue, a new cooperative jamming scheme is proposed in this paper, in which the data can be confidentially communicated from the source to the destination through multiple untrusted relays.  In our proposed secure transmission scheme, all the legitimate nodes contribute to providing secure communication by intelligently injecting artificial noises to the network in different communication phases. For the sake of analysis, 
we consider a multi-hop untrusted relaying network with two successive intermediate nodes, i.e, a three-hop communications network. Given this system model, a new closed-form expression is presented in the high signal-to-noise ratio (SNR) region for the Ergodic secrecy rate (ESR). Furthermore,
we evaluate the high SNR slope and
power offset of the ESR to gain an insightful comparison of the proposed secure transmission scheme and the state-of-arts. Our numerical results highlight that the proposed secure transmission scheme provides better secrecy rate performance compared with the two-hop untrusted relaying as well as the direct transmission schemes.
\end{abstract}

\vspace{5mm}
\section{Introduction}

{The open broadcast nature of wireless media, though makes communications ubiquitously accessible as the world has witnessed in the past decades, leads the security requirement to be a paramount challenge of such communications systems \cite{Zou2016WirelessSec}. Indeed, flow of large amount of data over wireless links that potentially may be sensitive in nature is more vulnerable than other transmission links to various security breaches such as location privacy \cite{Imran2017} and  eavesdropping attacks. Security in wireless communication networks is conventionally implemented above the physical layer using key-based cryptography methods {\cite {mukh}}. However that these computationally-based security methods have worked well in conventional systems, they may not be applicable to emerging fifth generation and beyond (B5G) wireless networks for the Internet of Things (IoT) which includes a broad range of applications such as unmanned aerial vehicle (UAV) networks, vehicular and ad-hoc networks (VANET), Internet of Vehicles (IoV), massive machine communication (MMC), and so forth \cite{Wang2019PLS5G}. As a matter of fact, these types of time-varying network topologies require complicated key management and sharing which is difficult to implement in distributed networks. Additionally, the computation and processing abilities of the nodes may be naturally limited and the complicated encryption calculations may not be supported \cite{Yener2015}. }

To complement these complex schemes, wireless transmitters can also be validated at the physical layer by exploiting the dynamic characteristics of the associated communication links {\cite {review1}}. {To accomplish this idea, physical layer security (PLS) has been emerged as a promising paradigm and an unbreakable security approach from information-theoretic perspective, and provisioned for safeguarding 5G wireless communications networks against eavesdropping attacks without incurring additional security overhead \cite {review1, IoT2019}. The fundamental notion of PLS is to intelligently exploit the characteristics of wireless channels and their impairments, e.g, noise, fading, diversity, etc, and possibly the information source \cite{Hamamreh2018PLS}. Indeed, the main design goal of PLS is to establish a performance gap between the link of the legitimate receiver, also known as \textit{Main link} and that of the eavesdropper, or the so called \textit{Wiretap link}, by using some well-designed secure transmission techniques (see, e.g., \cite{Sun2018review} and references therein).}

{In the context of PLS, cooperative jamming, which involves transmission of some  artificial noise signals to degrade the quality of received signal-to-noise ratio (SNR) at the potential eavesdropper while maintaining that at the intended destination, has been contemplated a powerful security technique \cite{mukh}. Further, cooperative jamming transmissions can be applied by any legitimate node of the network such as source \cite{Yang2015AN}, wherein some artificial noise is transmitted alongside the information signal, some dedicated authorized nodes \cite{Wang2015Jmrsel, Mamaghani2020UavJmr}, wherein extra helper entities are employed for the jamming transmission, or even intended receivers, which is named as {\it destination-assisted jamming} technique \cite{Mamaghani2019UAV, Kuhestani-IoT, Mamaghani2017sec}. Based on this technique, the decoding of the jammed signal by the authenticated user (destination) at some appropriate rate is possible, owning to the fact that the destination is able to receive quite a clean signal after self-interference cancellation, whereas the eavesdropper is kept unable to distinguish information-bearing signal from jamming transmitted by the destination.}

{Recently, the relay-assisted communications  wherein low-cost intermediate nodes may be exploited to assist the source-to-destination transmission, has attracted the attention of many researchers. Indeed, relaying technology can assist in providing reliable communication between the long-distance users, and improve the spectral efficiency and coverage.} Additionally, it has been viewed as a pervasive technology for the wireless sensor networks (WSN) and 5G IoT networks on the grounds that it can be adopted over a wide domain of applications such as smart homes, health-care services, device-to-device (D2D), IoV, and UAV communications \cite{IoT2019, Mamaghani2019UAV, UAV1, UAV2, Abro2019}. {For example, the authors in \cite{Mamaghani2019UAV} explored the end-to-end communications between long-distance users via UAV-assisted relaying. The authors in \cite{Abro2019} have considered a multi-hop scenario for a wireless sensor network and then  applied a Genetic algorithm  based optimization to enhance the energy expenditures, scalability and lifetime of the WSN.  Further, \cite{Hasan2018} presented a novel  interference mitigation algorithm with low complexity to improve the throughput and the outage performances for  Heterogeneous Networks (HetNets).
The problem of designing the optimum beamforming vector for multiuser multiple input multiple output (MU-MIMO) wireless communication system  to minimise interference has also been investigated in \cite{Adam2018} where the authors analyzed the Ergodic sum-rate capacity with Ricean fading channels. However, it should be mentioned that the aforementioned research works have considered the end-to-end communications  via trusted intermediate nodes, while the security issues of intermediate nodes have not been taken into account. }

{Nonetheless, several practical scenarios mentioned above may include \textit{untrusted relay} nodes, i.e., 
the intermediate nodes which lack perfect security clearance, from which the source-destination pair wishes to keep the confidential information to be exchanged secret in spite of enlisting their helps for the purpose of reliable communications \cite{IoT2019, sensors}. Hence, in these networks, it is important to protect the confidentiality of information from the untrustworthy relay, while simultaneously exploiting its relaying capability to improve the data transmission rate. }In the area of  untrusted relaying, an obvious yet thought-provoking question might initially arise  is whether or not a chain of untrusted relays can be beneficial for secure transmission of source-destination pair? 

\subsection{{Related works}}
{In the past decade, several works have considered the interesting scenario of untrusted relaying \cite{He, sun, Mamaghani2018sec, Mamaghani2018iet, Kuhestani-TCOM, Zhang2019UntrustedUAV, MH1, MH2, MH3}.
Thanks to the destination-based jamming strategy {\cite {He}}, it is shown that a positive secrecy rate can still be attained in untrusted relay networks. The authors is \cite{Kuhestani-TCOM} proposed a joint relay selection and power allocation scheme for an untrusted relaying scenario in the presence of either non-colluding passive eavesdroppers or colluding ones.
We note that non-colluding eavesdroppers independently try to obtain the confidential information without cooperating with each other. However, colluding eavesdroppers can potentially pose more harmful attacks by cooperatively attempting in decoding the confidential messages which is common in large-scale distributed networks \cite{Mirmohseni}.} { Further, Mamaghani \textit{et al.}, by performing secrecy performance analyses, thoroughly investigated a two-way secure untrusted amplify-and-forward (AF) relaying in the presence of an extra jammer \cite{Mamaghani2017sec, Mamaghani2018sec, Mamaghani2018iet}. In their system model, the energy-limited intermediate nodes are powered via simultaneous wireless information and power transfer (SWIPT) technique to establish a self-reliant secure relaying network. }

{
Notably most of the recent works \cite{sun, Mamaghani2018sec, Mamaghani2018iet, Kuhestani-TCOM} have focused on the simple scenario of two-hop untrusted relaying. In some communication networks such as ad-hoc and IoT, it is of great interest to  develop a communication network with more than two hops to provide the source-to-destination communication \cite{MH1, MH2, MH3}. Note that the consecutive relays may be necessary helpers to deliver the information signal to the destination. This is particularly valid when the communications channels experience a heavy shadowing where the communication environment gets harsh, or when the distance between terminals is large, or even  when the nodes suffer from limited power resources. Of the real-world communication systems that developing multiple relays might be crucial for message transmission are WSN, 5G IoT communications, and UAV-based relaying. For the latter application as an example, the wireless communication links may be easily blocked due to mountains and terrains in rural areas, or high-rise buildings in a metropolitan urban, which usually demands employing multiple UAV-relays to resolve the blockage or  long-distance issues \cite{Zhang2019UntrustedUAV}.}

Having said that, extending the analysis from two-hop to multi-hop untrusted relaying networks is non-trivial, because using more hops means  that more nodes get involved in the transmission as well as more chances for eavesdropping. In addition, the number of hops becomes a design parameter which affects on the end-to-end delay and throughput.  {Interestingly, He {\it et. al} in \cite{MH3} demonstrated that a non-negative secrecy rate can be achieved for  such a network by properly exploiting friendly jamming transmission from the appropriate nodes including both untrusted relays and destination. The main research question of that paper was whether an achievable non-vanishing perfect secrecy rate  is attainable regardless of how many unauthenticated intermediate nodes are required for establishing source-destination communication?  They showed that by employing an intelligent combination of some coding schemes, this goal could be achieved. It is worth pointing out that the untrusted relays considered in \cite{MH3} adopt compute-and-forward (CF) protocol for data transmission. However, in 5G IoT wireless networks which devices are low-power with limited computational capability, the designers aim at implementing architectures with low computational complexity, e.g., \cite{Xiangweng2018}. With that in mind, taking into account the CF relaying protocol, as investigated in \cite{MH3}, might become costly owning to the fact that each intermediate node needs to reliably retrieve the message from the received signal and then re-transmit what decoded to the next subsequent node.  Therefore, this relaying scheme may not be suitable for the computationally-limited intermediate nodes to be employed for end-to-end transmission protocol. As such, to the best of the authors' knowledge, the problem of low-complex untrusted relaying with more than two hops has not yet been extensively addressed.}

\subsection{{Our contribution}}
{
In this paper, considering the need for multiple relays for reliable communications, differing from \cite{sun, Mamaghani2018sec, Mamaghani2018iet, Kuhestani-TCOM}, and motivated by the solid work \cite{MH3}, we take into account secure transmission in a multi-hop AF untrusted relaying network. In the considered system model, a chain of intermediate nodes with low computational capabilities is necessary to facilitate end-to-end communications but a potential eavesdropper resides at each of them posing eavesdropping attack. Note that in AF relaying, the intermediate nodes simply forwards what they have received without attempting in decoding the information signal. Therefore, compared to CF relaying \cite{MH3}, AF relaying enjoys  more simplex structure.}

{The main contribution of this research work is threefold summarized as:}
\begin{itemize}

\item
{We propose a new  artificial noise injection based secure transmission protocol in a multi-hop {\it amplify-and-forward} untrusted relaying network
to keep the communication confidential from the internal eavesdroppers for any number of hops.}



\item
{We derive a novel closed-form expression for the ESR of three-hop untrusted relaying as a special case of multi-hop communications with two successive untrusted relays, at the high SNR. Furthermore, we characterize the high-SNR slope corresponding to the maximum multiplexing gain of the network, and the high-SNR power offset metric to obtain an insightful comparison of the proposed secure transmission scheme and the other benchmarks.
}

\item 
{We validate the theoretical analysis by comparing them with Monte-Carlo simulations, and demonstrate the significant secrecy performance improvement of the proposed cooperative jamming based multi-hop relaying over the conventional competitive benchmarks. We further study impacts of some key system parameters such as nodes distance, environmental path-loss, and transmission power on the overall system performance. Finally, we formulate a simple optimization problem to see how power allocation strategy could improve the secrecy performance of the system.}
\end{itemize}

{The rest of this paper is organized as follows. System model is given in Section 2, 
followed by detailing the proposed multi-hop transmission scheme with two relays and then deriving SNR representation at all the nodes in Section 3. Section 4 is dedicated  for the secrecy performance analysis of the proposed secure protocol where we derive new closed-form expressions for the ESR, as well as analyze the asymptotic
SNR metrics including the high SNR slope and the high SNR power offset. Next, numerical results and discussions are provided in Section 5 to illustrate the performance of the secure transmission scheme and obtain some key design insight into the proposed system model. Finally, conclusions are drawn in Section 6.}

\section{{System Model}}

\begin{figure}[htp]
	\centering
	\includegraphics[width=\columnwidth]{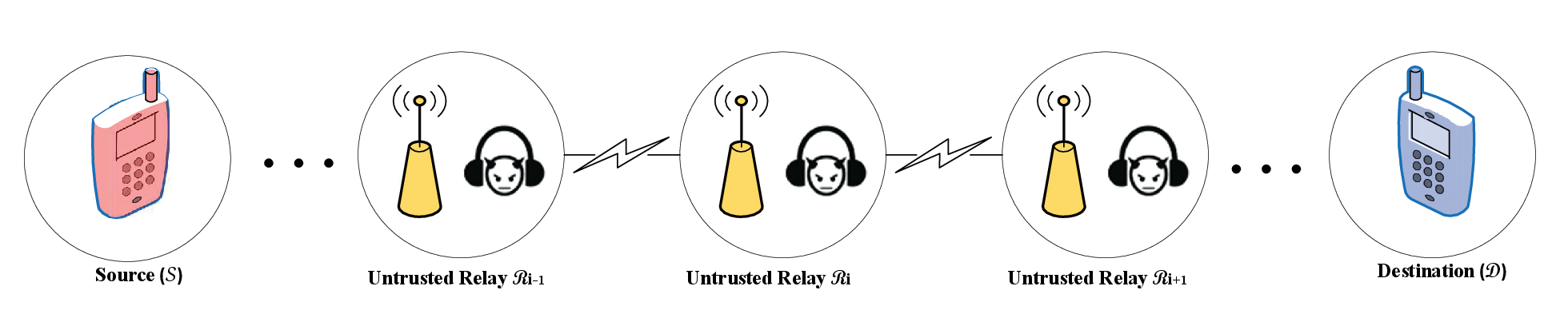}
	\caption{\small {System model of secure multi-hop untrusted relaying.}}
	\label{fig1:systemModel}
\end{figure}
{We propose a secure communications system via multi-hop untrusted relaying, as illustrated in Fig. \ref{fig1:systemModel},  where the source node, denoted by (\src), sends the information signal to the destination (\des) with the help of multiple consecutive relays, denoted by $\mathcal{R}_i$ where $i = 1, 2, \cdots N$. We  assume all the involving nodes are equipped with a single antenna operating in half-duplex mode, i.e., sending and receiving can not be done concurrently. In the so-called line network, it is also assumed that each node $\mathcal{R}_i$ can communicate with  its two neighbours $\mathcal{R}_{i-1}$ and $\mathcal{R}_{i+1}$ on the grounds that the channel quality between non-consecutive nodes is too weak to establish communications. Further, the AF relaying architecture is assumed operating at the relays where the relay simply forwards what they have received and so is very inexpensive to implement. Besides that, the intermediate nodes are assumed to be untrustworthy and hence, they would overhear the  transmitted information signal while relaying. Moreover, we assume that the relays are non-colluding, as for the line network taken into account it is less likely that non-consecutive intermediate illegitimate nodes could share their information and collude with each other due to channel conditions of these nodes. To be specific, it is assumed that the untrusted relays, at which the eavesdroppers residing, adopt selection combining (SC) processing technique to extract the information solely based on their own findings similar to \cite{sun, Kuhestani-TCOM}. Additionally, the channel  between any two consecutive nodes is assumed to follow channel reciprocity obeying complex Gaussian distribution\footnote{Note that this is a valid assumption for terrestrial networks, however, our work can be readily extended to consider other channel modelling based on the applications of interest, such as UAV-ground based channels as considered in \cite{Mamaghani2019UAV}}. Now, we turn our focus to the proposed secure transmission protocol and provide a detailed explanation with analysis for the considered system model when the number of relays is two, i.e., three-hop untrusted relaying, and then discuss the system performance for the higher number of relays, from engineering perspectives.}

\section{Transmission protocol}

\begin{figure}[htp]
	\centering
	\includegraphics[width=0.85\columnwidth]{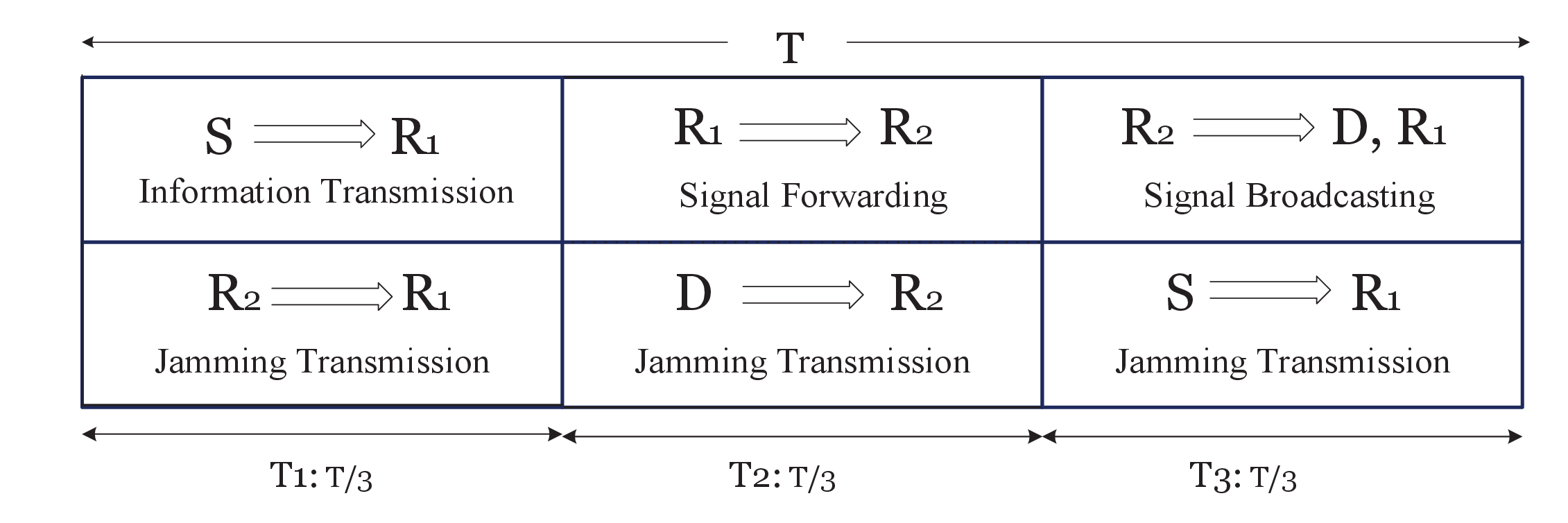}
	\caption{\small{ Secure transmission protocol of three-hop untrusted relaying.}}
	\label{fig2:DiagramProtocol}
\end{figure}

The proposed multi-hop secure untrusted relaying with two relays, i.e.,~\ra~\\and~\rb, can be detailed as follows. Considering a time division multiple access (TDMA) scheme, wherein the communication link is divided into separate time slots while sharing the same frequency band, one frame of transmission from~\src~to~\des~takes place in three phases, lasting $\frac{T}{3}$ seconds each, as shown in Fig. \ref{fig2:DiagramProtocol}.  At the outset, during the first phase of communication,~\src~transmits the information signal to~\ra~using superposition coding, and simultaneously,~\rb~jams the first untrusted relay (\ra) by transmitting an artificial noise. During the next phase,~\ra~forwards a scaled version of the received signal towards~\rb. Concurrently,~\des~jams~\rb~via transmitting a jamming signal  to guarantee secrecy. Finally, in the third phase,~\rb~amplifies and broadcasts the received signal which can be further received by~\des~and~\ra. 
After self-interface cancellation at~\des, the information signal can be extracted at~\des. Notably, during the last time slot, due to the fact that~\ra~can overhear the broadcasted signal by~\rb, the node~\src~is forced to emit a jamming to enhance the confidentially of communication. As such,~\ra~may fail to successfully eavesdrop during the last phase, as well. { Note that depending on the level of security required,~\src~ might be idle in the last phase, and therefore, an appropriate power allocation scheme plays an important role in the proposed cooperative jamming based untrusted relaying.}

Now, let assume that the complex Gaussian channel gains from \src~to~\ra,~\ra~to~\rb, and~\rb~to~\des\\are denoted by $g\sim \mathcal{CN}(0,m_g)$, $h\sim \mathcal{CN}( 0,m_h)$ and $f\sim \mathcal{CN}( 0,m_f)$, respectively. We here consider block fading channel model such that the channel coefficients vary independently from one frame to another frame, but do not change within one frame.   To make the analysis tractable, we consider the equal transmit power $P$ by the nodes, i.e., $P_s = P_{r_1}=P_{r_2}=P_d = P$. We also define $\gamma_g\treq\rho|g|^2$, $\gamma_h\treq\rho|h|^2$, and $\gamma_f\treq\rho|f|^2$, where $\rho=\frac{P}{N_0}$ describes the transmit SNR per each node. Remarkably $\gamma_g$, $\gamma_h$ and $\gamma_f$ follow  exponential distributions with means $\bar{\gamma}_g=\rho m_g$, $\bar{\gamma}_h=\rho m_h$, and $\bar{\gamma}_f=\rho m_f$, respectively.  Without loss of generality, the power of additive white noise at each receiver is considered to be $N_0$. We also suppose that the nodes are aware from the necessary channel state information (CSI), by which the relays as well as the destination can thoroughly cancel the self-interference term from the received signal. Note that this assumption leads to the maximum probability of eavesdropping at the relays, and in some sense, can be considered as the worst case scenario.

Based on the above descriptions and after some mass manipulations, the exact signal-to-interference-plus-noise-ratios (SINRs) at~\ra~in the first phase, at~\rb~in the second phase, at~\ra~and~\des~in the third phase, are respectively, obtained as 

\begin{equation}\label{gammar11}
\gamma^{(1)}_{R_1}=\frac{\gamma_g}{\gamma_h+1},
\end{equation}
\begin{align}\label{gammar22}
  \gamma^{(2)}_{R_2}=\frac{\gamma_g\gamma_h}{\gamma_g\gamma_f+\gamma_h\gamma_f+2\gamma_h+\gamma_g+\gamma_f+1},  
\end{align}
\begin{equation}\label{gammar13}
\gamma^{(3)}_{R_1}=\frac{\gamma_g\gamma_h^2}{\gamma^2_h+\gamma_h(\gamma_g+1)^2+(\gamma_g+\gamma_h+1)^2(\gamma_f+1)},
\end{equation}
\begin{equation}\label{gammadd}
\gamma^{(3)}_{D}=\frac{\gamma_g\gamma_h\gamma_f}{3\gamma_h\gamma_f+2\gamma_f\gamma_g+      \gamma_g\gamma_h+2\gamma_f+2\gamma_h+\gamma_g+1},
\end{equation}
where superscripts represent the phase of transmission. {Under the high SNR regime with $\gamma_k\gg1$ for $k\in \{g, h, f\}$, the above exact SINRs derived can be respectively, simplified as}
\begin{equation}\label{gammar11h}
\gamma^{(1)}_{R_1}\approx\frac{\gamma_g}{\gamma_h},~~~
\gamma^{(2)}_{R_2}\approx\frac{\gamma_g\gamma_h}{\gamma_f(\gamma_g+\gamma_h)},~~~
\gamma^{(3)}_{R_1}\approx\frac{\gamma_g\gamma_h^2}{(\gamma_g+\gamma_h)^2\gamma_f+2\gamma_h\gamma_g^2},
\end{equation}
\begin{equation}\label{gammaddh}
\gamma^{(3)}_{D}\approx\frac{\gamma_g\gamma_h\gamma_f}{3\gamma_h\gamma_f+2\gamma_f\gamma_g+\gamma_g\gamma_h}.
\end{equation}

{\bf Remark 1}: Expressions in (\ref{gammar11h}) reveal that the amount of information leakage is saturated when the transmit SNR goes to infinity. However, the received SINR at the legitimate receiver is a monotonically increasing function on the transmit SNR. As a result, the achievable ESR is increased as the transmit SNR grows which is fundamentally different from the direct transmission scheme \cite{Kuhestani-IoT}.\\

{\bf Remark 2}:  {As can be understood, in the proposed scheme, when a node transmits the information signal in the line of  destination, the node which is near to the receiving untrusted relay is forced to propagate artificial noise to confuse the eavesdropping node. As a consequence, this proposed scheme can be routinely extended to multi-hop untrusted relaying where more than two untrusted relays cooperate to forward a confidential message to the destination. Specifically, when $\mathcal{R}_{i-1}$ forwards the received signal to $\mathcal{R}_{i}$,~$\mathcal{R}_{i+1}$ jams the eavesdropper $\mathcal{R}_i$, and likewise when re-transmitting the amplified version of the received signal by~$\mathcal{R}_{i+1}$ to the next node, the nearest node to the transmitting relay, i.e., $\mathcal{R}_{i}$ may pose an eavesdropping attack, therefore, the relay  $\mathcal{R}_{i-1}$ is scheduled to send artificial jamming signals to make the wiretap link of $\mathcal{R}_{i}$ degraded. Setting $\mathcal{R}_0$ and $\mathcal{R}_{N+1}$ to be~\src~and~\des, respectively, the above explanation, can be readily extended for the multi-hop untrusted relaying. However, this extension gets too involved to analyze, and more importantly, may not be efficient for large number of hops. Since the latency of network grows by increasing the number of intermediate nodes \cite{sensors}. This is not acceptable in real-time communication scenarios. Furthermore, in IoT networks with low-cost and low-power equipment, the need of large overhead for the training process is challenging, especially when the environment is dynamic and thus, the coherence time of the wireless channel is short. As a result, in this work the three-hop untrusted scenario is considered for the purpose of analysis, though, may not be the optimal choice in terms of the number of relays. However, as we shall see later, the proposed multi-hop untrusted relaying with two relays improves the system performance.}\\

{\bf Remark 3}:  {
It is worth pointing out that in the considered line network which the end-to-end message delivering is conducted via multiple untrusted relays, we need to have careful scheduling and synchronization of transmissions based on the proposed TDMA-based protocol. Hence, all the network nodes are assumed to perform their transmissions in the equally-allocated time slots. Practically speaking, the communication channel, which could be considered as a sub-carrier of orthogonal frequency-division multiplexing
(OFDM) system,  is divided into separate time slots, and then, the communication process is accomplished through the mentioned multiple-phase protocol. Note that in this work, we have not considered scheduling and synchronization errors, each of which is worthy of deep investigation.}

\section{Secrecy Performance Analysis}
 Note that the  Ergodic secrecy rate (ESR), as a widely used secrecy criteria in the literature, characterizes the rate below which any average secure transmission rate can be obtained. 
In this section, we proceed to derive a new closed-form expression for the ESR of three-hop untrusted relaying. Based on the definition, the instantaneous secrecy rate is achieved by subtracting  the eavesdropping channel capacity from the legitimate channel capacity \cite{mukh}. Since the untrusted relays are non-colluding and they adopt SC technique, the instantaneous secrecy rate, $R_{s}$, for a three-hop relaying can be evaluated by
\begin{equation}\label{rsec1}
R_{s}=\frac{1}{3}\left[I^{(3)}_{D}-\max\{I^{(1)}_{R_1}, I^{(2)}_{R_2}, I^{(3)}_{R_1}\}\right]^+,
\end{equation}
where $	I_{K}=\log_{2}(1+\gamma_{K})$ with $K\in $ \{\ra,~\rb,~\des\} and $[x]^+ = \max(x,0)$. Notably the pre-log factor  $\frac{1}{3}$ is due to the fact that one round of transmission is done in three hops.

{\bf Remark 3}:
It is worth noting that $\gamma^{(2)}_{R_2}\gg\gamma^{(3)}_{R_1}$, which can be readily concluded by invoking $\gamma^{(2)}_{R_2}$ and $\gamma^{(3)}_{R_1}$ expressions given in (\ref{gammar11h}). Therefore, the maximum information leakage of three-hop untrusted relaying can be simplified as
\begin{align}\label{gammaE}
    \gamma_E \treq \max\left\{\gamma^{(1)}_{R_1}, \gamma^{(2)}_{R_2}\right\},
\end{align}

{The exact ESR expression of the proposed three-hop untrusted relaying can be obtained by forming a multiple integral expression as
\begin{align}\label{asr}
    	\bar{R}_s&= \E\{R_s\}=\int_{0}^{\infty}\int_{0}^{\infty}\int_{0}^{\infty}R_s(p,q,r)f_{\gamma_g}(p)f_{\gamma_h}(q)f_{\gamma_f}(r)dp dq dr,
\end{align}
wherein $R_s(p,q,r)$ is given in \eqref{rsec1}, and the probability density functions (PDFs) $f_{\gamma_t}(s)=\frac{1}{\bar{\gamma}_t}\exp\left(-\frac{s}{\bar{\gamma}_t}\right)$ with $s\geq 0$ for $t\in $ \{g, h, f\}. Although, the integral expression given above can be calculated numerically, in order to obtain deep insight into the impact of parameters in secrecy rate, we are interested in obtaining a new compact expression for the ESR. To that aim, we first derive closed-form expressions for the  Ergodic legitimate rate $\bar{R}_L$ and the  Ergodic eavesdropping rate $\bar{R}_E$ in the following lemmas, and thereafter, we will be ready to present a tight lower-bound expression for the ESR performance in Proposition \ref{prop1}.}

\begin{lemma}
	The lower-bound closed-form expression for the Ergodic rate of the legitimate channel, without pre-log factor, is given by
	\begin{align}\label{ER}
	\bar{R}_L&=\E\Big\{\log_2(1+\gamma_D)\Big\}
	\nonumber\\
	&\geq  \log_2\left(1+\exp\left[-3\Phi+\ln\left(\frac{\bar{\gamma}_g\bar{\gamma}_h\bar{\gamma}_f}{3\bar{\gamma}_h\bar{\gamma}_f+2\bar{\gamma}_f\bar{\gamma}_g+\bar{\gamma}_g\bar{\gamma}_h}\right)\right]\right) \stackrel{\Delta}{=} \bar{R}_L^{LB},
	\end{align}
	where $\Phi \approx 0.577215$ is the Euler constant.
\end{lemma}
{\it Proof:} The proof can be done straightforwardly by considering the facts that: 1) the Jensen's inequality can apply on the convex function $\ln(1+\exp{(x)})$ with respect to $x$ and, 2) for the exponential random variable (RV) $X$ with the mean of $m_X$, we have $\E\{\ln (X)\}=-\Phi+\ln (m_x)$ {\cite[Eq. (4.331.1)] {integ}}.

Before proceeding further to derive an analytical expression for $\bar{R}_E$, we present the following fruitful Lemma.
\begin{lemma}\label{lemma1}
	Let $X$ and $Y$ be exponential RVs with means $m_x$ and $m_y$, respectively. Then the new RVs $Z=\frac{X}{Y}$ and $W=\frac{XY}{X+Y}$ have the following cumulative distribution functions (CDFs), respectively, as
	\begin{align}\label{fw}
	F_Z(z)&=\frac{m_yz}{m_yz+m_x},\\
	F_W(w)&=1-\frac{2\omega}{\sqrt{m_x m_y}}\exp\left(-\frac{\omega}{m_x}-\frac{\omega}{m_y}\right)\mathrm{K}_1\hspace{-1mm}\left(\frac{2\omega}{\sqrt{m_xm_y}}\right),
	\end{align}
	where	$\mathrm{K}_\nu(\cdot)$ is the modified Bessel function of the second kind and $\nu$-th order.
\end{lemma}
{\it Proof:} See Appendix A.

\begin{lemma} \label{lemma2}
	The closed-form expression for the  Ergodic rate of the eavesdropping channel, without pre-log factor, is formulated as 
	\begin{align}\label{EvR}
	\bar{R}_E&=\E\Big\{\log_2(1+\gamma_E)\Big\}\nonumber\\
	&=\frac{1}{\ln 2}\E\left\{\ln\left(1+\mathbbm{1}_{\left\{\gamma^{(1)}_{R_1}\geq \gamma^{(2)}_{R_2}\right\}}\gamma^{(1)}_{R_1} + \mathbbm{1}_{\left\{\gamma^{(2)}_{R_2}> \gamma^{(1)}_{R_1}\right\}}\gamma^{(2)}_{R_2} \right)\right\}\nonumber\\
	&=\frac{1}{\ln 2}\E\left\{\mathbbm{1}_{\left\{\gamma^{(1)}_{R_1}\geq \gamma^{(2)}_{R_2}\right\}}\ln\left(1+ \gamma^{(1)}_{R_1} \right)+\mathbbm{1}_{\left\{\gamma^{(2)}_{R_2}> \gamma^{(1)}_{R_1}\right\}}\ln\left(1+ \gamma^{(2)}_{R_2} \right)\right\}\nonumber\\
	&=\frac{1}{\ln 2}\Big(\mathcal{P} T_{1}+(1-\mathcal{P})T_2\Big),
	\end{align}
	where $\mathbbm{1}_{\left\{X\right\}} = \begin{cases}
         1  & \text{if}~~~\text{{X} = True} \\
             0  &  \qquad~~~ \text{o.w.},
       \end{cases}$~ represents the indicator function which is one iff its condition is satisfied, and the last equation follows from total probability theorem or simply considering the expectation of indicator function which can be further calculated analytically as
	\begin{align}
	\mathcal{P}&=\E\left\{\mathbbm{1}_{\left\{\gamma^{(1)}_{R_1}\geq \gamma^{(2)}_{R_2}\right\}}\right\} =\Pr\left\{\gamma^{(1)}_{R_1}\geq \gamma^{(2)}_{R_2}\right\}\nonumber\\
	&={\frac {\sqrt {\bar{\gamma}_f}{\bar{\gamma}_g}^{3/2}}{\bar{\gamma}_g-\bar{\gamma}_h\,\sqrt {\bar{\gamma}_f}\sqrt {\bar{\gamma}_g}+2\,\bar{\gamma}_g\,\bar{\gamma}_h}}\nonumber\\
	&\times
	\sum_{n=1}^{M}\sum_{i=1}^{n}\Lambda(1,n,i)i!\left({\frac {2\bar{\gamma}_g\,\bar{\gamma}_h}{\bar{\gamma}_g-\bar{\gamma}_h\,\sqrt {\bar{\gamma}_f\,\bar{\gamma}_g}+2\,\bar{\gamma}_g\,\bar{\gamma}_h}}
	\right)^i,
	\end{align}
	where the parameter $M$ holds an arbitrary positive integer value controlling the approximation accuracy. Also, 
	\[\Lambda(\nu, n, i)= \frac{(-1)^i\sqrt{\pi}\Gamma(2\nu)\Gamma(n-\nu+\frac{1}{2})\mathrm{L}(n,i)}{2^{\nu-i}\Gamma(\frac{1}{2}-\nu)\Gamma(n+\nu+\frac{1}{2})n!},\] where $\mathrm{L}(i,n)={{{n-1\choose i-1}}\frac{n!}{i!}}$ for $n, i >0$ represents the Lah numbers \cite{Molu2017}, $\Gamma(x)$ is the Gamma function {\cite[Eq. (8.31)] {integ}}. Furthermore,
	\begin{align}
	T_{1}=\frac{\bar{\gamma}_g\ln(\frac{\bar{\gamma}_g}{\bar{\gamma}_h})}{\bar{\gamma}_g-\bar{\gamma}_h},
	\end{align}
	\begin{align}
	T_{2}=\ln\left(1+\frac{\bar{\gamma}_g \bar{\gamma}_h \Big(\bar{\gamma}_g^2-\bar{\gamma}_h^2-2\bar{\gamma}_g\bar{\gamma}_h\ln\frac{\bar{\gamma}_g}{\bar{\gamma}_h}\Big)}{	\bar{\gamma}_f
(\bar{\gamma}_g-\bar{\gamma}_h)^3}\right),
	\end{align}
	
	\begin{proof}
		See Appendix B. 
	\end{proof}
\end{lemma}
\begin{proposition}\label{prop1}
	The tight closed-form lower-bound expression for the ESR performance of the proposed three-hop untrusted relaying is given by
	\begin{eqnarray}\label{R_LB}
	\bar{R}^{LB}_{s}=\frac{1}{3}\Big[\bar{R}_L^{LB}-\bar{R}_E\Big]^+.
	\end{eqnarray} 
\end{proposition}
\subsection{Asymptotic Ergodic Secrecy Rate Analysis}
Now , we are going to obtain the asymptotic expression for the ESR performance, denoted by $ R^{\infty}_s$, when the transmit SNR of each node, $\rho$, goes to infinity by deriving the high SNR slope $S_{\infty}$ in bits/s/Hz and the high SNR power offset $L_{\infty}$ in 3dB unit. These parameters are defined, respectively, as \cite{Kuhestani-TCOM}
\begin{equation}\label{SL}
S_{\infty}=\lim_{\rho\to\infty}\frac{\bar{R}_{s}}{\log_{2}\rho}~~\mathrm{and}~~ L_{\infty}=\lim_{\rho\to\infty}\big({\log_{2}\rho}-\frac{\bar{R}_{s}}{S_{\infty}}\big),
\end{equation}

\begin{align}\label{Rs:asympt}
    R^{\infty}_s = S_{\infty} \left( \log_2 \rho - L_{\infty}\right),
\end{align}
Following the same steps
as in \cite{Mamaghani2018sec}, the high SNR slope and power offset of the three-hop untrusted relaying are obtained as
\begin{eqnarray}\label{slope}
S_{\infty}=\frac{1}{3},
\end{eqnarray}
and
\begin{align}\label{offset}
L_{\infty}&=\frac{3\Phi}{\ln 2} +\log_2\left(\frac{3}{m_g}+\frac{2}{m_h}+\frac{1}{m_f}\right)
+ \mathcal{P}_1\mathcal{A} +(1-\mathcal{P}_1)~\mathcal{B},
\end{align}
where 
\begin{align}
\mathcal{A}&=\frac{m_g(\log_2({m_g})-\log_2({m_h}))}{m_g-m_h},\\
     \mathcal{B}&=\log_2\left[1+\frac{m_g m_h \left( m_g + m_h -2m_h\mathcal{A}\right)}{	m_f
(m_g-m_h)^2}\right],
 \end{align}
{{\it Proof:} The analytical expressions of $S_{\infty}$ and $L_{\infty}$ can be readily obtained by plugging \eqref{prop1}, while considering \textit{Remark 1}, into the definition of high-SNR slope and power offset given by \eqref{SL} and using the approximation $\log(1+x)\approx \log(x)$ when $x\gg1$. Besides, $\mathcal{P}_1$ is defined in Appendix A. Hence, we skipped the details for the sake of brevity.}
 
Expression (\ref{slope}) highlights that the channel powers have
no impact on the ESR slope which is equal to the maximum multiplexing gain of the network. Furthermore, different from the high SNR slope, we find that the high SNR power offset in (\ref{offset}) is related to the all channel powers. As such, by properly positioning the relays between the source and destination, the high SNR power offset can be reduced. Notably, a decrease in the power offset corresponds to an increase
in the ESR performance.

\section{Numerical Results and Discussions}
In this section, we prepare some simulations to reveal the accuracy of the presented closed-form expressions. Additionally, we compare the secrecy performance of the proposed multi-hop relaying scheme with two competitive counterparts: 1) the two-hop  communication scheme where only one relay is selected for data transmission and the other relay is considered as pure eavesdropper, and 2) the direct transmission where the confidential information is directly forwarded to the destination without getting assistance from the two relays. In this case, both the relays are considered as pure eavesdroppers. Unless otherwise stated, the following simulation parameters are adopted. For simplicity and without loss of generality,
we assume that the nodes~\src,~\des,~\ra,~and~\rb~are placed on one-dimensional space at positions $-3$, $+3$, $-1$ and $+1$, respectively. 
Additionally, the large-scale path-loss factor is chosen $\eta$= 2.7. Besides, for Mont-Carlo simulations we averaged over $10^5$ channel realizations.

\begin{figure}[t]
	\centering  
	\includegraphics[width=0.85\columnwidth]{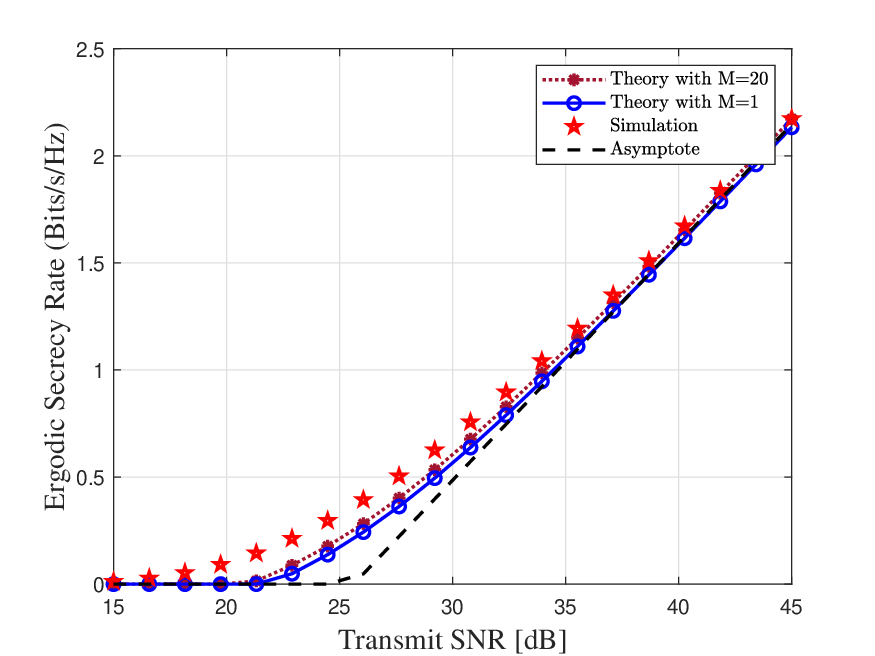}
	\caption{{Validation of theoretical results for the ESR performance.}}
	\label{fig2:Validation}
\end{figure}

Fig. \ref{fig2:Validation} is plotted to depict the ESR performance versus the transmit SNR $\rho$ (in dB). As can be seen in this figure, our proposed lower-bound expression for the ESR in Proposition 1 agrees well with the exact ESR. Furthermore, our asymptotic ESR performance given by \eqref{Rs:asympt} well-approximates the exact ESR in the high SNR regime. As observed from Fig. \ref{fig2:Validation}, the ESR curve corresponding to the case when considering only the first term of the equivalent series, denoted by \textit{Theory with $M=1$} is so close to the exact ESR curve, especially in high SNR regime. This reaffirms the accuracy and tightness of analytical expressions we obtained.

To reveal the advantage of the proposed three-hop untrusted relaying scheme, we compare the ESR performance of our new scheme with two well-known transmission schemes, i.e., two-hop untrusted relaying and direct transmission. {Note that under two-hop relaying scheme, we face with two cases. In Case I, the relay~\ra~is employed for data re-transmission and the relay~\rb~is considered as a pure eavesdropper, as illustrated in Fig. 4(a). Whereas, in Case II, the converse scenario is considered, i.e., the relay~\rb~is the helper node and~\ra~is considered as an idle eavesdropper,  as illustrated in Fig. 4(b).} Additionally, two network topologies are considered. In Topology 1, we have the same network structure as considered in Fig. \ref{fig2:Validation}, and for Topology 2, we have the scaled version of Topology 1 with factor of $3$, i.e., the nodes~\src,~\des,~\ra~and~\rb~are located at -9, +9, -3 and +3, respectively. 

\begin{figure}[t]

\subfloat[Case I: Relaying using~\ra,~while~\rb~is considered as an idle adversary.]{%
  \includegraphics[clip, width=0.85\columnwidth]{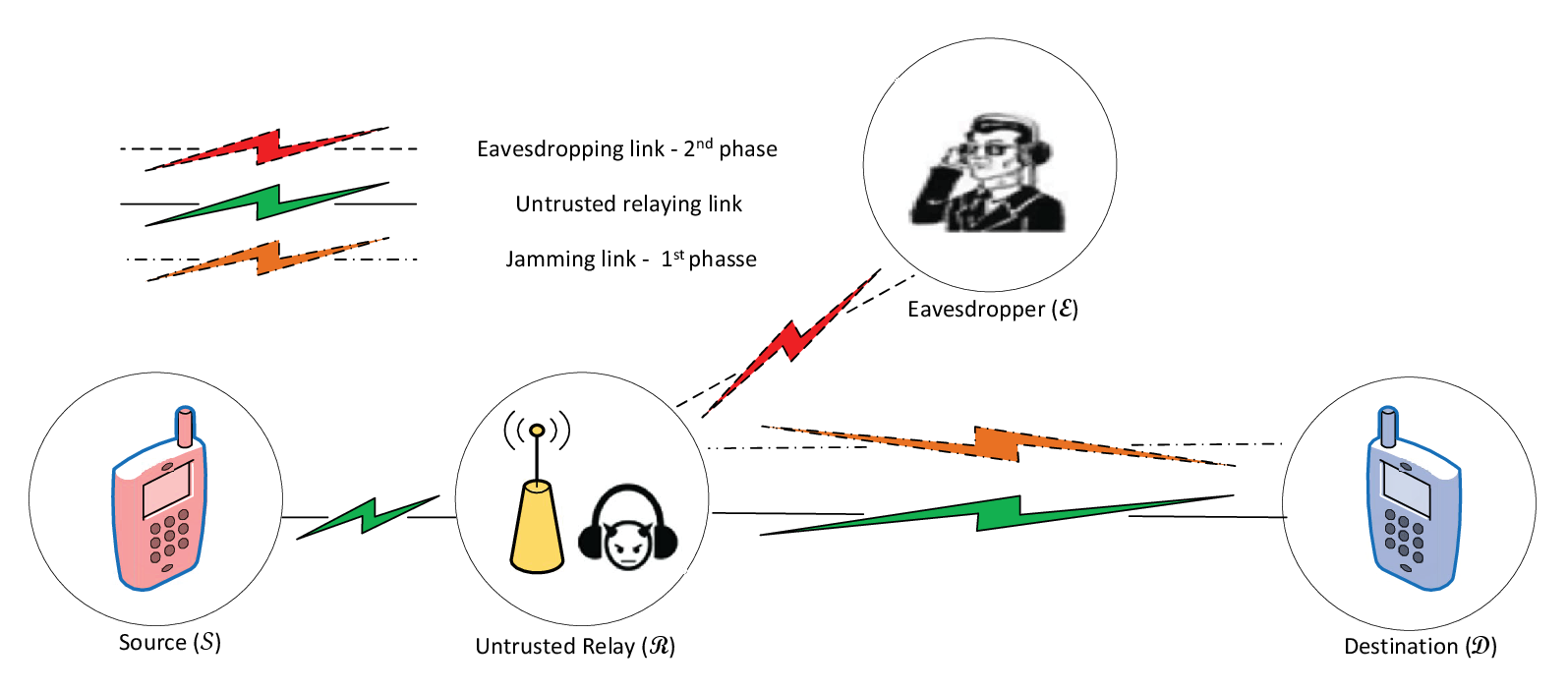}
}

\subfloat[Case II: Relaying using~\rb,~while~\ra~is considered as an idle adversary.]{%
  \includegraphics[clip, width=0.85\columnwidth]{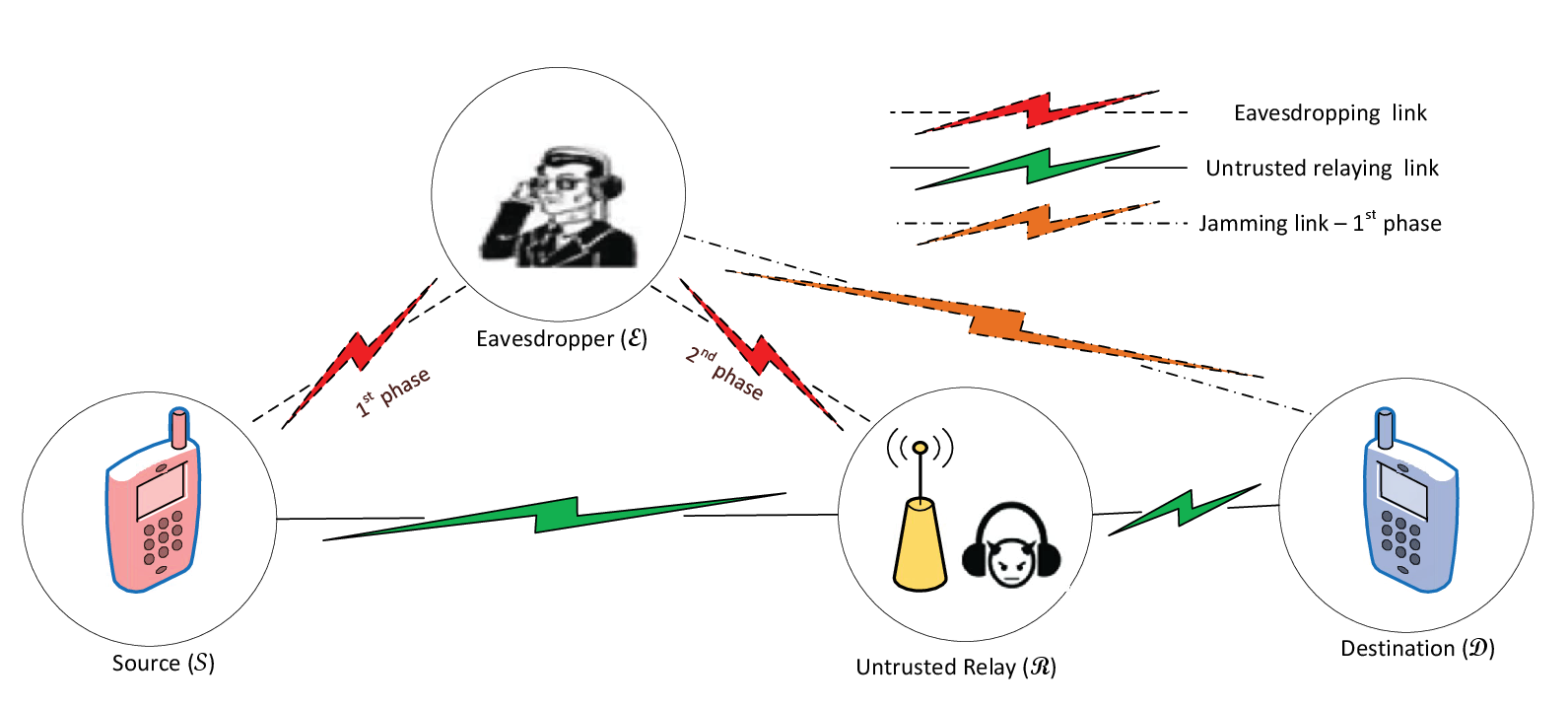}  
}
\caption{{Considered two-hop untrustworthy relaying as benchmarks.}}

\end{figure}


\begin{figure}[t]
	\centering  
	\includegraphics[width = 0.85\columnwidth]{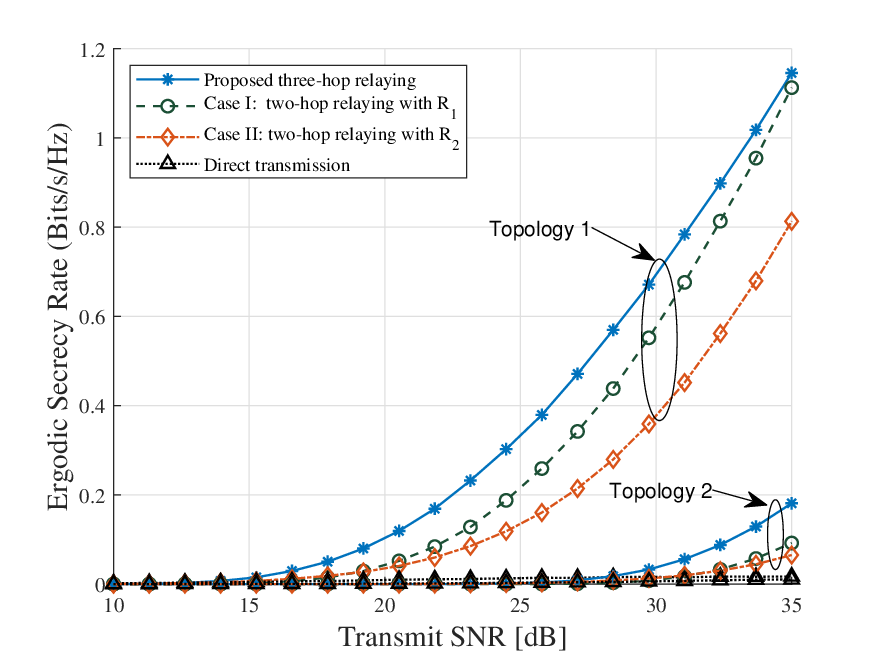}
	\caption{{ESR versus transmit SNR for different transmission schemes and topologies.}}
	\label{fig3:Comparison}
\end{figure}

 As it can be observed in Fig. \ref{fig3:Comparison}, the secrecy performance of the proposed three-hope relaying scheme always outperforms the  mentioned two benchmarks for the transmit SNRs fewer than approximately 36 dB (i.e., $\rho <$ 36 dB). This result highlights the priority of our scheme compared with the state-of-the-arts in untrusted relaying networks. One can easily predict that the proposed scheme under Topology 2 outperforms the two-hop relaying schemes for $\rho>$ 36 dB. Interestingly, under Topology 1 and for $\rho>$ 36 dB, the two-hop relaying scheme with Case I would provide a better ESR compared with our scheme. The reason is that when the communication nodes are close together with much power budget, naturally, the two-hop relaying would be sufficient for data transmission and hence, it is not necessary to implement multi-hop relaying scheme. Additionally, as proved in \cite{Kuhestani-TCOM}, the high SNR slope for two-hop relaying is $S_\infty=\frac{1}{2}$  which is more than the high SNR slope of tree-hop relaying scheme, $S_\infty=\frac{1}{3}$, as we derived in (\ref{slope}). It is worth noting that in IoT and WSNs, the devices are power-limited and thus, they cannot consume much power for data transmission/forwarding. As a result, the proposed secure three-hop relaying scheme in this paper is applicable for IoT networks where low or medium transmit SNRs can be supported by the devices. Finally, this figure depicts that the direct transmission scheme presents a near to zero, but non-zero, secrecy rate. As discussed in \cite{Kuhestani-TCOM}, even when the destination is very far from the source while the eavesdroppers  locate between them, a positive secrecy rate is achievable.

\begin{figure}[t]
	\centering  
	\includegraphics[width = 0.85\columnwidth]{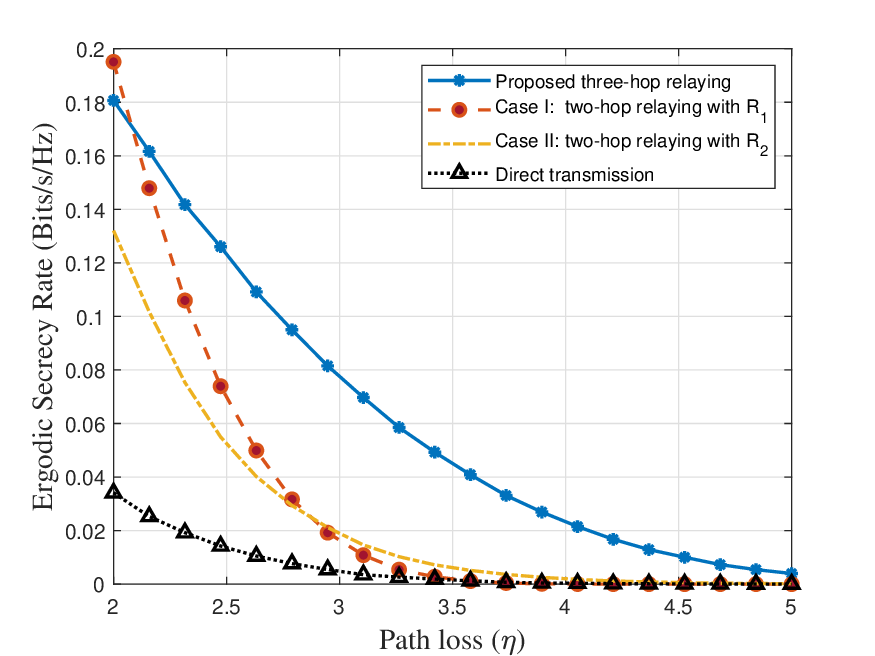}
	\caption{{ESR versus path-loss exponent for different transmission schemes. SNR is set to $\rho = 20$dB.}}
	\label{fig4:Comparison}
\end{figure}

{Fig. \ref{fig4:Comparison} exhibits the ESR performances of the proposed multi-hop untrusted relaying and the known schemes against the path-loss exponent. Obviously, the ESR gets decreased for higher order of path-loss values which demonstrates severe fading or blockage results in lower ESR. Nonetheless, the proposed scheme with two relays provides a significantly pronounced ESR performance for practical values of $\eta$. For example, the ESR of the proposed scheme has threefold improvement compared to the two-hop relaying for path-loss of 3. This again boosts the effectiveness of our proposed secure scheme in real-world applications such as WSN and low altitude UAV-based relaying network \cite{Mamaghani2019UAV} wherein fading or blockage are undeniable.}

\begin{figure}[t]
	\centering  
	\includegraphics[width = 0.85\columnwidth]{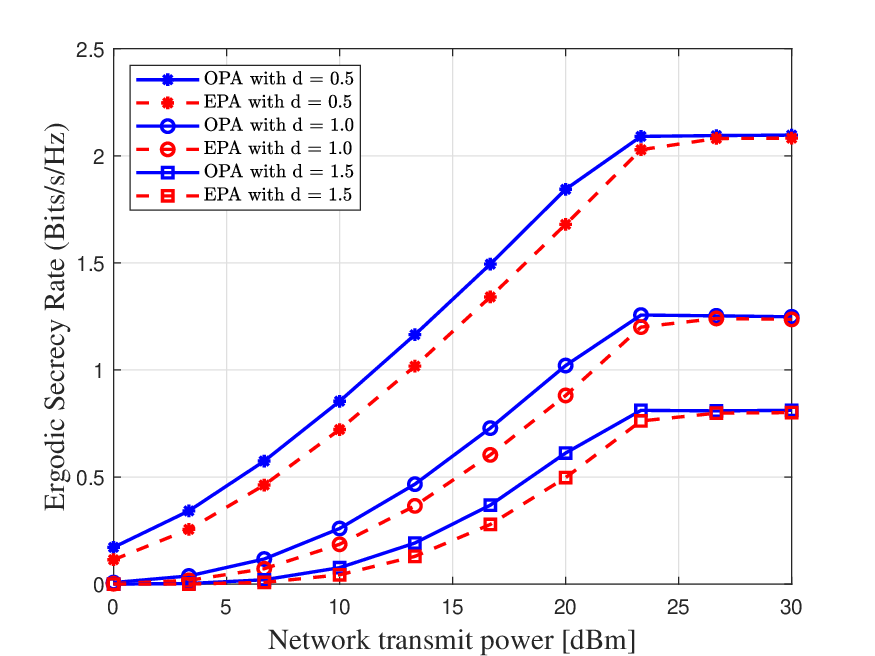}
	\caption{{ESR versus network power budget.}}
	\label{fig6:NetPow}
\end{figure}

\begin{figure}[t]
	\centering  
	\includegraphics[width = 0.85\columnwidth]{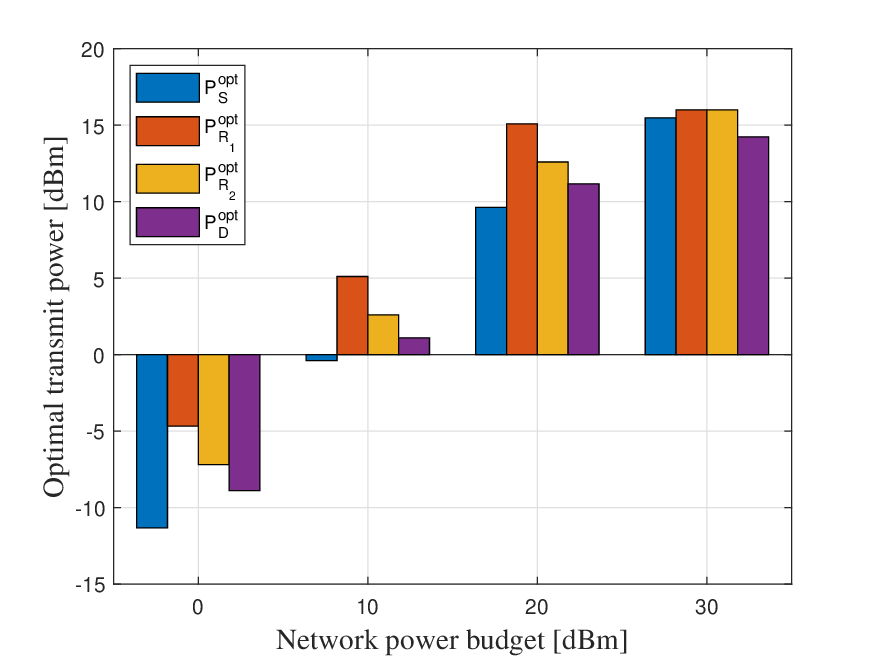}
	\caption{{OPA versus network power budget.}}
	\label{fig5:OPAdis}
\end{figure}

{Now, in order to obtain a better understanding of the proposed system, Figs. \ref{fig6:NetPow} and \ref{fig5:OPAdis} are provided 
to demonstrate how optimizing the network resources, i.e., allotted power to each network node, could bring improvement in terms of ESR performance. Specifically, in  Fig. \ref{fig6:NetPow} the optimal power allocation (OPA) curves are obtained via solving the optimization problem formulated as follows
\begin{align}\label{OPA}
\hspace{-1mm}(P):&\stackrel{}{\underset{{{P_s, P_{r_1}, P_{r_2}, P_{d}}}}{\mathrm{maximize}}~~~~~\frac{1}{N}\sum_{n=1}^{N} R^n_s({P_s, P_{r_1}, P_{r_2}, P_{d}})} \nonumber\\
&~~~\text{subject to}~~~~~ \mathrm{C1}:~0 \leq P_k \leq P_{max},\nonumber\\
&~~~~~~~~~~~~~~~~~~~~~\mathrm{C2}:~\sum_k P_k \leq P_{tot},~~\text{for}~k\in\{s, r_1, r_2, d\}
\end{align}
where the objective function is averaging over $N$ realization of the channels which is approximately equal to the the expected value of $R_s$ given by \eqref{asr}. We mention that as $N$ goes to infinity, the approximation turns into equality based on the law of large numbers \cite{papoulis}. In \eqref{OPA}, the constraint $C1$  represents the transmission power of each node subject to the maximum power $P_{max}$, and $C2$ is the constraint for network power budget $P_{tot}$ defined as $P_{tot} = 2P_s+P_{r_1}+2P_{r_2}+P_d$ based on the proposed transmission protocol.  Note that the above problem is non-convex due to non-smooth and non-concave objective function. Additionally, owning to the fact that performance improvement does not fall into the main scope of this research work, we just  use the optimization toolbox of Matlab R2020a to solve the above problem sub-optimally. The following system parameters are considered for simulation: noise power $N_0 = -20$ dBm,
$P_{max} = 16$ dBm, and $N=10^5$ channel realizations.  The equal power allocation (EPA) curves are obtained by setting equal transmission power per phase per active node, i.e., $P_s=P_{r_1}=P_{r_2}=P_d= \frac{P_{tot}}{6}$. This is due to this fact that one frame of transmission is conducted in three phases and two nodes are transmitting in each phase of communications.
As it can be clearly found from Figs. \ref{fig6:NetPow} and \ref{fig5:OPAdis}, the least proportion of power budget should be allocated to~\src~for low-to-moderate range of network power budget, whilst~\ra~should enjoy a higher transmission power compared to the other nodes, in order to get the ESR improved.}

\section{Conclusions}
{In this contribution, we designed a new secure transmission scheme over multi-hop untrusted relaying networks. To this end, we first studied a three-hop communication network with two successive untrusted relays. Given this system model, a novel closed-form expression was derived in the high SNR regime for the ESR performance. We
next evaluated the high SNR slope and
power offset of the ESR. 
Our numerical results presented that the proposed secure transmission scheme  improves the secrecy performance compared with the competitive benchmarks, i.e., the two-hop relaying and the conventional direct transmission schemes. As a future work, we could consider the impacts of imperfect CSI and hardware impairments on the secrecy performance of the considered multi-hop untrusted relaying as well as resource allocation problem for of such a network.}

\appendix

\section{Proof of Lemma 2}
The CDF of $Z=\frac{X}{Y}$ has been derived in \cite{Kuhestani-IoT}. To obtain the CDF of $W=\frac{XY}{X+Y}$, we start from the definition of CDF as\\
\begin{align}
F_W(\omega)&=\Pr\Big\{\frac{XY}{X+Y}<\omega\Big\}=\Pr\Big\{XY-\omega(X+Y)<0\Big\}\nonumber\\
&=\Pr\Big\{X<\frac{\omega Y}{Y-\omega}|Y-\omega\geq 0\Big\}\Pr\{Y-\omega\geq 0\}\nonumber\\
&+{{\Pr\Big\{X\geq\frac{\omega Y}{Y-\omega}|Y-\omega<0\Big\}}}\Pr\{Y-\omega<0\}\nonumber\\&=\int_{\omega}^{\infty}F_X\left(\frac{\omega y}{y-\omega}\right)f_Y(y)dy+\int_{0}^{\omega}f_Y(y)dy\nonumber\\
&=\int_{\omega}^{\infty}\hspace{-1.5mm}\left[1-\exp\left(-\frac{\omega y}{m_x(y-\omega)}\right)\right]f_Y(y)dy+\int_{0}^{\omega}f_Y(y)dy\nonumber\\
&=1-\frac{1}{m_y}\int_{\omega}^{\infty}\exp\left(-\frac{\omega y}{m_x(y-\omega)}-\frac{y}{m_y}\right)dy\nonumber\\
&=1-\frac{1}{m_y}\exp\left(-\frac{\omega}{m_x}-\frac{\omega}{m_y}\right)\int_{0}^{\infty}
\exp\left(-\frac{\omega^2}{m_x y}-\frac{y}{m_y}\right)dy\nonumber\\
&\stackrel{(a)}{=}1-\frac{2\omega}{\sqrt{m_x m_y}}\exp\left(-\frac{\omega}{m_x}-\frac{\omega}{m_y}\right)\mathrm{K}_1\hspace{-1mm}\left(\frac{2\omega}{\sqrt{m_xm_y}}\right),
\end{align}
Finally, after calculating the integral term using {\cite[Eq. (3.324.1)] {integ}}, one can obtain the expression given in (\ref{fw}).
\done

\section{Proof of Lemma 3}

In the following, we proceed to prove Lemma 3 wherein the different exact/approximate expressions for $\mathcal{P}$, $\mathcal{T}_1$, and $\mathcal{T}_2$ are given.

{\bf {A. Calculating} $\mathcal{P}$:} Plugging (\ref{gammar11h}) into $\mathcal{P}=\Pr\{\gamma^{(1)}_{R_1}>\gamma^{(2)}_{R_2}\}$, and then defining $X=\gamma_f$, $Y=\gamma_h$ and $Z=\gamma_g$, we get 
\begin{align}\label{P}
\mathcal{P}&=\Pr\left\{\gamma_f>\frac{\gamma^2_h}{\gamma_g+\gamma_h}\right\}=1-\Pr\left\{X<\frac{Y^2}{Y+Z}\right\}\nonumber=1-\E_Y\left\{\E_Z\left\{F_X\left(\frac{y^2}{y+z}\right)\right\}\right\}\nonumber\\
&=\frac{1}{m_ym_z}\int_{0}^{\infty}\int_{0}^{\infty}\exp\left(-\frac{y^2}{(y+z)m_x}-\frac{y}{m_y}-\frac{z}{m_z}\right)dzdy\nonumber\\
&\stackrel{(a)}{=}\frac{1}{m_ym_z}\int_{0}^{\infty}\int_{0}^{\infty}\exp\left(-\frac{v^2}{um_x}-\frac{v}{m_y}-\frac{u-v}{m_z}\right)dudv\nonumber\\
&\stackrel{(b)}{=}\sqrt{\frac{4}{m_xm_ym_z}}\int_{0}^{\infty}\hspace{-3mm}\exp\left(-v\frac{m_y-m_z}{m_ym_z}\right)v{{K}_1}\left(\sqrt{\frac{2}{{m_xm_z}}v}\right)dv\nonumber\\
&\stackrel{(c)}{\approx}{\frac {\sqrt {m_{x}}{m_{z}}^{3/2}}{m_{z}-m_{y}\,\sqrt {m_{x}}\sqrt {m
			_{z}}+2\,m_{z}\,m_{{\it y}}}}\sum_{n=1}^{M}\sum_{i=1}^{n}\Lambda(1,n,i)i!\left({\frac {2m_{z}\,m_{y}}{m_{z}-m_{y}\,\sqrt {m_{x}\,m_{z}}+2\,m_{z}\,m
		_{y}}}
\right)^i,\nonumber\\
&\stackrel{(d)}{\approx}{\frac {4\sqrt {m_{x}}{m_{z}}^{5/2}m_{y}}{3 \left(m_{z}-m_{y}\,\sqrt {m_{x}}\sqrt {m_{z}}+2\,m_{z}\,m_{y} \right)^{2}}}\treq\mathcal{P}_1,
\end{align}\\
where $(a)$ follows from defining the auxiliary variables $u=y+z$ and $v=y$, $(b)$ follows 
from using {\cite[Eq. (3.324.1)] {integ}} and {\cite[Eq. (3.351.3)] {integ}}, $(c)$ follows from 
substituting the  equivalent series of modified Bessel function of the second kind and first order as presented in \cite{Molu2017}, which is a well-tight approximation with finite series, as observed later in numerical results. For $\nu>0$ and positive integer $M$ which controls the accuracy of infinite series, we have \cite{Molu2017} 
\[
{\sl K}_{\nu}(\beta x)\approx\exp(-\beta x)\sum\limits_{n=0}^{M}\sum_{i=0}^{n}\Lambda(\nu, n, i)(\beta x)^{i-\nu},\]
Finally, $(d)$ presents the first term of the infinite series given for $M=1$ to have a closed-form approximation. We will show in the simulation results  how this simple closed-form expression works well.

\noindent{\bf {B. Calculating}} $T_1$:
Using Lemma \ref{lemma1}, we can derive a closed-form expression for $T_{1}$, after assuming $X=\frac{\gamma_g}{\gamma_h}$, as 
\begin{eqnarray}
T_{1}=\E\left\{\ln(1+\frac{\gamma_g}{\gamma_h})\right\}=\int_{0}^{\infty}\ln(1+x)f_X(x)~dx \stackrel{(a)}{=}\int_{0}^{\infty}\frac{1-F_X(x)}{1+x}dx,
\end{eqnarray}
where $(a)$ follows from integration by parts law. Then, computing the last integral, considering $F_X(x)$ given in Lemma \ref{lemma1}, leads to the closed-form expression for $T_1$ given in (\ref{t22}).\\
{\bf {C. Calculating}} $T_2$:
The part $T_{2}$ can be mathematically calculated as
\begin{eqnarray}\label{t22}
\hspace{-3mm}T_{2}=\E\left\{\ln\left(1+\frac{\gamma_g\gamma_h}{\gamma_f(\gamma_g+\gamma_h)}\right)\right\}\stackrel{(a)}{\approx}\ln\left(1+\frac{\E\left\{\frac{\gamma_g\gamma_h}{\gamma_g+\gamma_h}\right\}}{\E\{\gamma_f\}}\right),
\end{eqnarray}
where $(a)$ follows after using the approximation $\E\left\{\log\left(1+\frac{X}{Y}\right)\right\}\approx\log\left(1+\frac{\E\{X\}}{\E\{Y\}}\right)$ given in \cite{Bjornson}. Thus, after further calculation, using Lemma 2, one can obtain $T_2$ in (\ref{t22}).

\done



\end{document}